\documentclass[pra, twocolumn, amsmath, amssymb]{revtex4}
\usepackage{graphics,hyperref}

\begin{document}
\date{\today}
\title{Non-Markovian Open Quantum Systems}
\author{C\'{e}sar A. Rodr\'{i}guez-Rosario} 
\email[email: ]{carod@physics.utexas.edu}
\affiliation{  The University of Texas at Austin, Center for Complex Quantum Systems, 1 University Station C1602, Austin TX 78712}
\author{E.~C.~G. Sudarshan}
\affiliation{  The University of Texas at Austin, Center for Complex Quantum Systems, 1 University Station C1602, Austin TX 78712}

\begin{abstract}
We construct a non-Markovian canonical dynamical map that accounts for systems correlated with the environment. The physical meaning of not completely positive maps is studied to obtain a theory of non-Markovian quantum dynamics. The relationship between inverse maps and correlations with the environment is established. A generalized non-Markovian master equation is derived from the canonical dynamical map that goes beyond the Kossakowski-Lindblad Markovian master equation. Non-equilibrium quantum thermodynamics can be be studied within this theory.
\end{abstract}


\maketitle 

\section{Introduction}

The theory of open quantum systems was first introduced as the quantum analogue of classical stochastic processes \cite{Sudarshan61a}. The evolution of a system that interacts with outside degrees of freedom is fully given by a dynamical map that corresponds to a quantum stochastic process. The state, the environment and their correlations change with time. If the environment is assumed not to react on the system, the Markov approximation can be taken in which these correlations are discarded to derive the Kossakowski-Lindblad master equation \cite{Kossakowski72a,Gorini76a,Lindblad76a}. This theory extends quantum mechanics beyond Hamiltonian dynamics, and has been crucial to the study of quantum thermodynamics for phenomena such as decoherence \cite{Breuer02a}.

The Markov approximation is unreasonable for many physical phenomena. Even if the environment is large compared to the system, it might still react on the system for very short times. For short times, the system can only couple to a few environmental degrees of freedom. These will act as a memory. Short time scales in experiments often show environmental memory effects. For example, in the case of spin-echoes a decay can be partially undone by exploiting environmental memory effects \cite{Hahn50a}. Also, non-Markovian quantum effects may play a role in the energy transfer in photosynthesis  \cite{Engel07a}. 

Modeling non-Markovian open quantum systems is crucial for understanding these experiments. The Kossakowski-Lindblad master equation proves to be inadequate. Extensions to the theory of open quantum systems to go beyond the Markov approximation have been developed \cite{Shabani05a,Uchiyama06a,Breuer07a}, but the theory is incomplete. In this paper we develop a generalization of open quantum systems to states correlated with their environment, leading to non-Markovian dynamics.

We review the theory of stochastic processes for quantum systems in Section \ref{sectstoc}. Different forms of the dynamical maps, their inverses and properties are discussed. To highlights the limitations of the Kossakowski-Lindblad master equation, we review the conditions and approximations necessary to derive it from the dynamical map are described in Section \ref{kossak}. In Section \ref{sectnonmarko} we study how initial correlations naturally limit the domain of valid physical states and not completely positive dynamical maps can arise. We construct a canonical dynamical map is constructed that is non-Markovian, with the understanding that the compatibility domain of not completely positive dynamical maps is connected to correlations of the system with the environment. In Section \ref{sectmaster} we derive the generalized non-Markovian master equation. The equation is local in time. A canonical embedding map dynamically determines how the system is correlated to the environment at all times. In Section \ref{sectnoneq} we introduce irreversibility by discarding terms of higher order in time. Truncation does not eliminate all the memory effects of the environment and correspond to non-equilibrium quantum thermodynamical phenomena. We discuss the connection of the non-Markovian master equation to previous instances of specific non-Markovian master equations and make the concluding remarks in Section \ref{sectconclu}.

The theory of non-Markovian open quantum systems we developed in this paper provides a simple way to describe decoherence phenomena beyond the Kossakowski-Lindblad master equation.

\section{Quantum Stochastic Processes}\label{sectstoc}

The density matrix $\rho$ describes the most general quantum mechanical state. The density matrix must have unit-trace, Hermiticity, and non-negative eigenvalues \cite{vonNeu}. These are related to the properties of classical probability vectors and allow us to interpret the expectation values of density matrices as physical observables. If we write the density matrices using tensor notation, a quantum stochastic  process acts like a classical stochastic process, as described in Appendix \ref{classtoc}. A quantum stochastic supermatrix $\mathbb{A}$ can be defined to describe the most general evolution of an initial density matrix $\rho(i)$ to a final density matrix $\rho(f)$,
\begin{equation}\label{AMap} 
\rho(i)_{rs}\rightarrow \rho(f)_{r^\prime s^\prime}= \mathbb{A}_{r^\prime s^\prime,rs} \;\rho(i)_{rs}.\nonumber
\end{equation}
The supermatrix $\mathbb{A}$ acts on the density matrix $\rho$ like if it was a vector,
$\overrightarrow{\rho}(f) =\mathbb{A}\cdot \overrightarrow{\rho}(i)$.
The quantum stochastic supermatrix has the properties: 
\begin{subequations}\label{AProperties} 
\begin{align}
\mathbb{A}_{r^\prime r^\prime,r s} & =\delta_{r,s}, \\
\mathbb{A}_{s^\prime r^\prime,s r} & =\mathbb{A}^*_{r^\prime s^\prime,r s}, \\
x^*_{r^\prime} x_{s^\prime} \mathbb{A}_{r^\prime s^\prime,r s}y_{r}y^*_{s} & \geq0. 
\end{align}
\end{subequations}
The first property guarantees the preservation of the trace, the second property preserves Hermiticity, while the last property imposes the condition that non-negative density matrices are mapped into non-negative density matrices and may be referred to as \emph{positivity} \cite{Sudarshan61a}. In Section \ref{sectinve}, we study physical situations where the positivity condition must be relaxed. 

We observe that in this form the composition of two maps, $\mathbb{A}^\prime \star \mathbb{A}$,  is the matrix multiplication of their supermatrices $\mathbb{A}^\prime_{r^\prime s^\prime,r^{\prime\prime} s^{\prime\prime}} \mathbb{A}_{r^{\prime\prime} s^{\prime\prime},rs}$. A pseudo-inverse $\widetilde{\mathbb{A}}$ can be defined such that:
\begin{equation}\label{ainver}
\widetilde{\mathbb{A}}_{r^\prime s^\prime,r^{\prime\prime} s^{\prime\prime}} \mathbb{A}_{r^{\prime\prime} s^{\prime\prime},rs}=\delta_{(r^\prime s^\prime),(rs)}.\end{equation}
The matrix $\widetilde{\mathbb{A}}$ is positive only on a convex domain consisting of a subset of all density matrices. Its action is only well-behaved on the subset $\{ \rho^\prime \}$ of density matrices of the form $\overrightarrow{\rho}^\prime =\mathbb{A}\cdot \overrightarrow{\rho}$ for all $\left\{ \overrightarrow{\rho} \right\} $. This subset is called the \emph{compatibility domain}. Outside the compatibility domain, the positivity of the density matrix need not be preserved by the inverse map. \footnote{ The compatibility domain was first introduced to describe system states compatible with initial correlations with their environment \cite{jordan:052110,jordan06a}.}

We simplify the properties of $\mathbb{A}$ by an index exchange of the form $\mathbb{B}_{r^\prime r,s^\prime s}\equiv\mathbb{A}_{r^\prime s^\prime,rs}$ and obtain: 
\begin{subequations}\label{BProperties} 
\begin{align}
\mathbb{B}_{r^\prime r,r^\prime s}  & =  \delta_{r,s},\\
\mathbb{B}_{r^\prime r,s^\prime s} & =   \mathbb{B}^*_{s^\prime s,r^\prime r},\\
x^*_{r^\prime} y^*_{r} \mathbb{B}_{r^\prime r,s^\prime s} x_{s^\prime} y_{s}   & \geq  0.
\end{align}
\end{subequations}
Hermiticity is now guaranteed because the map $\mathbb{B}$ itself is Hermitian. However, in this form the action of the superoperator $\mathbb{B}$ is not simply matrix multiplication on a vector $\overrightarrow{\rho}$ as it was in the $\mathbb{A}$ form. Instead, the map $\mathbb{B}$ acts in the following manner: 
\begin{eqnarray}\label{BMap} 
\rho(i)_{rs}\rightarrow \rho(f)_{r^\prime s^\prime}= \mathbb{B}_{r^\prime r,s^\prime s} \;\rho(i)_{rs},\nonumber
\end{eqnarray} or just $\rho(i)\rightarrow \rho(f)=\mathbb{B}\; \rho(i) $ for short. In this form, the composition of two maps $\mathbb{B}^\prime \star \mathbb{B}$ is $\mathbb{B}^\prime_{r^\prime r^{\prime\prime},s^\prime s^{\prime\prime}}\mathbb{B}_{r^{\prime\prime} r,s^{\prime\prime} s}$, which is \emph{not} matrix multiplication as it was in the $\mathbb{A}$ form. The inverse $\widetilde{\mathbb{B}}$ can be calculated from $\widetilde{\mathbb{A}}$ by exchanging the indices, inheriting its compatibility domain.

Now, we decompose $\mathbb{B}$ into its eigenmatrices and real eigenvalues, $\mathbb{B}_{r^\prime r,s^\prime s}\; \rho(i)_{rs}=\sum_{\alpha} \lambda(\alpha)\, \mathrm{C}(\alpha)_{r^\prime r} \,\rho(i)_{rs}\, \mathrm{C} ^*(\alpha)_{s s^\prime}$. The action of the map is:
 \begin{eqnarray}\label{CMap} 
\rho(f)\equiv \sum_\alpha\lambda_\alpha \,\mathrm{C}_\alpha\, \rho(i)\, \mathrm{C}^\dagger_\alpha.
\end{eqnarray} 
Note that $\{\mathrm{C}_\alpha\}$ are linearly independent and trace orthonormal, $\mbox{Tr}[\mathrm{C}^\dagger_\alpha \mathrm{C}_\beta]=0$ for $\alpha \neq \beta$. Hermiticity of $\rho$ is automatically preserved by the multiplication on the left and right. The trace of $\rho$ is preserved by the condition $\sum_\alpha \lambda_\alpha\mathrm{C}^\dagger_\alpha \mathrm{C}_\alpha=\openone$. The positivity condition is still only implicit. A stronger condition, \emph{complete positivity}, is very natural now. Complete positivity is defined as having all non-negative eigenvalues $\lambda_\alpha\geq0$ \cite{Choi72a,Choi75}. Complete positivity is a condition on the map itself, while positivity is a condition on the action of the map on density matrices. Much attention has been given to this class of maps, but confining quantum evolution to them has proven to be too restrictive \cite{pechukas94a,Buzek01a,jordan:052110,Terno05a,Rodriguez07a,ziman06}. Inverse maps are generally not completely positive \cite{Jordan07b}. In this paper we will use the $\mathbb{A}$ form, $\mathbb{B}$ form and its eigen-decomposition to show how  the complete positivity condition incompatible with non-Markovian open quantum systems.

\subsection{Dynamical Maps of Open Quantum Systems}

The evolution of a closed quantum system is generated by a unitary operator $ U_{(t_f|t_i )}=e^{ -i (t_f - t_i) H } $ \footnote{For compactness, we assume a time independent Hamiltonian $H$, but the results of this paper also apply to the time dependent case.}. The differential form of the evolution is given by the von Neumann equation, $\dot{\rho}(t)=-i\left[H,\rho(t)\right]$. The evolution can also be viewed as a stochastic process through a unitary map, 
\[\mathbb{U}_{(t_f|t_i )}\rho(t_i)\equiv U_{(t_f|t_i )} \rho(t_i) U^{\dagger}_{(t_f|t_i )}=\rho(t_f).\] This map is completely positive. The inverse of this map, $\widetilde{\mathbb{U}_{(t_i|t_f )}}\equiv\mathbb{U}_{(t_f|t_i )}^\dagger=\mathbb{U}_{(t_i|t_f )}$ has the whole set of density matrices as its compatibility domain, making it bi-stochastic. 

We are interested in the evolution of an open quantum system. In this case, the total state $\rho^{\mathcal{SE}}$ has a part that is accessible to us, the system $\mathcal{S}$, and one that is inaccessible, a finite-dimensional environment $\mathcal{E}$. The density matrix of the system space is found by tracing-out the environmental variables, $\eta^{\mathcal{S}}=\mbox{Tr}_{\mathcal{E}}\left[\rho^{\mathcal{SE}}\right]$ \footnote{From now on the total system-environment space will be denoted by $\rho$, while the reduced system state by $\eta$. Superscripts to indicate the system $\mathcal{S}$ and environment $\mathcal{E}$ will be suppressed when their meaning is clearly implied.}. If we only monitor the evolution of the system, it is generally non-unitary and best described by a dynamical map of the form:
\begin{equation}\label{BDynMap} 
\mathbb{B}_{(t_f|t_i )}\eta(t_i) \equiv\mbox{Tr}_{\mathcal{E}}\left[U_{(t_f|t_i )}
\rho(t_i) U^\dag_{(t_f|t_i )}\right]=\eta(t_f).
\end{equation}
In the full space, the evolution is given by the unitary map $\mathbb{U}$, while in the reduced space we get a more complicated evolution:  
\begin{eqnarray}\label{DynMapPicture} 
\rho(t_i) & \longleftrightarrow & \rho(t_f)=U_{(t_f|t_i )}\rho(t_i) U^\dag_{(t_f|t_i )} \nonumber \\
\downarrow & & \downarrow  \nonumber \\
\eta(t_i) &  \dashrightarrow  &\eta(t_f)=\mbox{Tr}_{\mathcal{E}}\left[\rho(t_f)\right]. 
\end{eqnarray}
The top level of the diagram represents the unitary evolution of the total system. The lower level is the reduced, open system, evolution. To go from the total space to the reduce space, or ``down" as indicated by the arrows, we use the trace map $\mathbb{T} \rho^{\mathcal{SE}}\equiv {\mbox{Tr}}_{\mathcal{E}} \rho^{\mathcal{SE}} = \eta$.

Note that there is no arrow to go from $\eta \rightarrow \rho$, or  ``up''. There is no map that inverts the trace such that $\widetilde{\mathbb{T}}\star\mathbb{T} \rho^{\mathcal{SE}}=\rho^{\mathcal{SE}}$. Inverting the trace would depend on a kernel that comes from correlations of the system with the environmental variables. In Section \ref{sectembed} we will derive a pseudo-inverse map that inverts the trace based on the physical considerations of the dynamics of the total state.  With such a map, the dynamical map for the process from $\eta(t_i)\rightarrow\eta(t_f)$ can be expressed as the composition of three maps,
\begin{eqnarray}\label{Compo}
\mathbb{B}_{(t_f|t_i )}\equiv \mathbb{T} \star \mathbb{U}_{(t_f|t_i )} \star\widetilde{\mathbb{T}}.
\end{eqnarray}
First, the trace is inverted to go from the reduced to the total space, then a unitary map evolves the total state and finally a trace reduces it to the system part of the space.
When $t_f=t_i$, there is no evolution and the map is just unity. How to properly define a pseudo-inverse map that connects the reduced space to the total space for all times is one of the main results of this paper. 

\subsection{Initially Uncorrelated States}\label{sectInitUnco}

A standard assumption for the evolution of an open system of the form Eq.~(\ref{BDynMap}) is that the system and environment are at the initial time in a Kronecker product of two density matrices, $\rho^\mathcal{SE}(t_i)= \eta^{\mathcal{S}}(t_i)\otimes\tau^{\mathcal{E}}$. Uncoupling a system from its environment may not be accomplished in experiments \cite{Kuah07a}. This very restrictive assumption can be shown to lead to dynamical maps that, in the form of  Eq.~(\ref{CMap},) have non-negative eigenvalues \cite{SudChaos}. This is proven by breaking the corresponding dynamical map into the composition of several completely positive maps, as in Eq.~(\ref{Compo}). The reduction at the end of the evolution $\mathbb{T}$ and the unitary map $\mathbb{U}$ are both completely positive. The map $\widetilde{\mathbb{T}}$ can be defined as an embedding map $\mathbb{E}$  \cite{pechukas94a,alicki95} that takes the system state at the initial time, and embeds it into the system-environment space:
\begin{eqnarray}\label{embedcp} 
\widetilde{\mathbb{T}}\Big(\eta(t_i)\Big) \equiv  \mathbb{E}_{t_i}\Big(\eta(t_i)\Big)=\eta(t_i)\otimes\tau. 
\end{eqnarray}
Since $\tau$ has positive eigenvalues, the embedding map can be written as $\mathbb{E}\left(\eta\right)= \big(\openone^{\mathcal{S}}\otimes \sqrt{\tau^{\mathcal{E}}} \big) \eta^{\mathcal{S}} \big(\openone^{\mathcal{S}}\otimes \sqrt{\tau^{\mathcal{E}}} \big)^\dagger$, which is of the form of Eq.~(\ref{CMap}) with non-negative eigenvalues. The dynamical map that takes a state without initial correlations with its environment is the composition of three completely positive maps: embedding, unitary evolution, and reduction. Initially uncorrelated states are not the only states that can give rise to completely positive maps \cite{Rodriguez07a}. 

The embedding map presented here is only applicable to the system at time $t_i$. At other times it might have developed correlations with the environment and not be of the product form. A generalization of this map for all times will be presented in Section \ref{sectembed}. 

\subsection{Example}\label{examp1}

To illustrate the relationship between the different forms of the map, we compute a simple example of a two-level system, a qubit, represented by the Bloch vector $\overrightarrow{a}$. Its most general transformation in the affine form \cite{jordan:034101} is:
\begin{eqnarray}
\overrightarrow{a}(t_f)=\overline{R}_{(t_f|t_i )}\cdot\overrightarrow{a}(t_i)+\overrightarrow{r},
\end{eqnarray}
 where the matrix $\overline{R}$ squeezes and rotates the Bloch vector, and the vector $\overrightarrow{r}$ translates. For this example, we focus on the particular case where the system interacts with a two-level uncorrelated environment $\tau=\frac{1}{2}\openone$. The total initial state is:
 \begin{eqnarray}\label{examptensor1}
\rho(t_0)^{\mathcal{SE}}=\frac{1}{2}\Big(\openone^{\mathcal{S}} + a_j(t_0) \sigma_j^{\mathcal{S}} \Big)\otimes\frac{1}{2}\openone^{\mathcal{E}},
\end{eqnarray}
where summation over the repeated index $j$ is implied, and $\sigma_j$ are the Pauli spin matrices. The system $\mathcal{S}$ is described by the Bloch vector $\overrightarrow{a}$. The environment at the initial time is fully mixed. If we assume a unitary operator that depends on the Hamiltonian $H=\sum_j\frac{1}{2}\sigma_j^\mathcal{S}\otimes\sigma_j^\mathcal{E}$, the evolution of the Bloch vector is:
\begin{eqnarray}\label{exampAffine1}
\overrightarrow{a}(t)=\cos\left(t-t_0\right)^2\overrightarrow{a}(t_0),
\end{eqnarray}
which is a uniform squeezing with no translation \cite{Rodriguez05a}. This interaction was chosen because it swaps the system with the environment at periodic intervals, thus storing the system information in the environment and then returning it. As time changes, the state is pinned to the fully mixed state and grows again into the full state periodically. 

The evolution can be treated as a map from $\eta(t_0)\rightarrow\eta(t)$ with the form from Eq.~(\ref{BDynMap}). If the density matrix $\eta(t)= \frac{1}{2}\left(\openone + a_j(t) \sigma_j \right)$ is written as a vector,
\begin{eqnarray}
\overrightarrow{\eta}(t)=\frac{1}{2}\left(%
\begin{array}{c}
  1+a_3(t) \\
  a_1(t)-i a_2(t) \\
  a_1(t)+i a_2(t) \\
  1+a_3(t) \\
\end{array}
\right),\nonumber
\end{eqnarray}
the evolution is a stochastic matrix transformation $\overrightarrow{\eta}(t)=\mathbb{A}_{(t|t_0 )}\cdot\overrightarrow{\eta}(t_0)$, where
\begin{eqnarray}
\mathbb{A}_{(t|t_0 )}=\frac{1}{2}\left(%
\begin{array}{cccc}
  1+c^2 & 0 & 0 & 1-c^2 \\
  0 & 2c^2 & 0 & 0 \\
  0 & 0 & 2c^2 & 0 \\
  1-c^2 & 0 & 0 & 1+c^2 \\
\end{array}%
\right) ,\nonumber
\end{eqnarray}with $c\equiv \cos\left(t-t_0\right)$. By index exchange, we get the map in its Hermitian form:
\begin{eqnarray}\label{exampBmap}
\mathbb{B}_{(t|t_0 )}=\frac{1}{2}\left(%
\begin{array}{cccc}
  1+c^2 & 0 & 0 & 2c^2 \\
  0 & 1-c^2 & 0 & 0 \\
  0 & 0 & 1-c^2 & 0 \\
  2c^2 & 0 & 0 & 1+c^2 \\
\end{array}%
\right).
\end{eqnarray} By rewriting the map in terms of its eigenvalues and eigenmatrices, $\eta(t)=\sum^3_{\alpha=0} \lambda_\alpha(t-t_0)\mathrm{C}_\alpha \eta(t_0) \mathrm{C}_\alpha^\dagger$ with

\begin{eqnarray}
 \lambda_0 (t-t_0)=\frac{1}{2}\left(1+3c^2 \right),& \; &\mathrm{C}_0=\frac{1}{\sqrt{2}}\openone,\nonumber\\
\lambda_{1,2,3} (t-t_0)=\frac{1}{2}\left(1-c^2 \right),&\; &\mathrm{C}_{1,2,3}=\frac{1}{\sqrt{2}}\sigma_{1,2,3},\nonumber
\end{eqnarray} 
we confirm that it is completely positive and trace preserving.

The process is reversible. Since the environment is finite dimensional, there are Poincar\'e recurrences. Also, note that even if this map is expanded in a Taylor series for $t\approx t_0$, where: 
\begin{eqnarray}
c^2=\cos\left(t-t_0\right)^2= 1-\left(t-t_0\right)^2+\ldots  ,\nonumber
\end{eqnarray} 
there are no terms of first order. To get irreversibility from this example, we will need to perform and approximation. In the next section, we review the approximations necessary to obtain the Kossakowski-Lindblad master equation.

\section{Kossakowski-Lindblad Master Equation}\label{kossak}

So far we have only discussed reduced Hamiltonian dynamics on quantum systems where no information is lost. In this section we discuss the Kossakowski-Linblad master equation in order to model monotonic decay of a quantum state. The master equation can be interpreted as the time-derivative of the dynamical maps after a series of assumptions and approximations.

 A dynamical map of a system that was uncorrelated from the environment at $t_0$ might developed correlations through time, its history reduces the allowed set of states at time $t_1$ such that,
\begin{eqnarray}\label{nocompo}
\mathbb{B}_{(t_2|t_0 )}\neq\mathbb{B}_{(t_2|t_1 )}\star \mathbb{B}_{(t_1|t_0 )}. 
\end{eqnarray}
The Markov approximation, assumed to be valid for short times, makes
\begin{eqnarray}\label{aproxcompo}\mathbb{B}_{(t_2|t_0 )} \approx\mathbb{B}_{(t_2|t_1 )}\star \mathbb{B}_{(t_1|t_0 )}.
\end{eqnarray} 
The maps now form a dynamical semigroup. This approximation was used by Kossakowski to derive a family of equations of motion for open quantum system
 \cite{Kossakowski72a,Gorini76a,Lindblad76a} that lead to irreversible behavior. The solutions to Kossakowski-Lindblad master equation often lead to exponential decays. Exponential decay solutions are a consequence of the approximations necessary to derive the master equation, and are connected to equilibrium quantum thermodynamics.
The deviation from exponential for short times is the quantum Zeno effect \cite{Misra77a}. 

We need to take the Markov approximation to obtain the approximated composition property from Eq. (\ref{aproxcompo}) in order to properly define the derivative of the dynamical map at all times. Also, we must assume that the system is initially uncorrelated from its environment, and thus the evolution is completely positive. In addition, we must perform a time rescaling to get the Kossakowski-Lindblad master equation from the dynamical map. With these, we obtain the equation:
\begin{eqnarray}\label{lindblad}
\frac{\partial\eta}{\partial t}&=&-i\left[H ,\eta\right]\nonumber\\&+&\sum_{\alpha}\frac{1}{2}\left(2\mathrm{L}_\alpha \eta \mathrm{L}^\dagger_\alpha -\mathrm{L}^\dagger_\alpha \mathrm{L}_\alpha\eta  -\eta \mathrm{L}^\dagger_\alpha  \mathrm{L}_\alpha\right),
\end{eqnarray}
The details of this derivation are given in Appendix \ref{dkossak}.

Altogether, these restrictions can describe dissipative processes at the expense of discarding non-Markovian memory effects and correlations with the environment. The Markov approximation can lead to unphysical results \cite{Jordan07a}. States with initial environmental correlations can have not completely positive dynamics \cite{Shaji05a}. The discarded higher orders of $t$ introduced irreversibility into the equation \cite{Lindblad83}, giving rise to thermodynamic effects. The  destruction of the memory effects make the present independent of the past and the state independent from its correlations with the environment \cite{Spohn80a}. These limit the physical situations where the Markov approximation can be applied.  We question each these assumptions.

\subsection{Example}

To illustrate the approximations made to obtain the Kossakowski-Lindblad master equation we study an example. We derive a thermodynamic decay from the collision model developed by Rau \cite{Rau63a}. Consider the evolution that leads to Eq. (\ref{exampAffine1}). We can model decoherence by treating the total environment as a stream of $\left\{\tau_i\right\}$, where each of them interact sequentially for a short average time $T$. This corresponds to acting with the dynamical map from Eq. (\ref{exampBmap}) in sequence: 
\begin{equation} \label{eq:Decoh} \eta^{\mathcal{S}}\rightarrow \mathbb{B}_{\left(t_{n},t_{n-1}\right)}\star\mathbb{B}_{\left(t_{n-1},t_{n-2}\right)}\star\ldots\star\mathbb{B}_{\left(t_1,t_0\right)}\left(\eta^{\mathcal{S}}\right),\nonumber\end{equation} 
where each time interval has duration $T=t_m-t_{m-1}$. After $N$ interactions the total time $t=NT$ has passed and the density matrix has the form:
\begin{equation}
\eta(t)=\frac{1}{2}\Big(\mathbb{I}^{\mathcal{S}} + \cos\left(T\right)^{2N} a_j (t_0)\sigma_j^{\mathcal{S}} \Big). \nonumber
\end{equation}
The shrinking factor can be rewritten as \[\cos\left(T\right)^{2N}=e^{-t\frac{2}{T}\ln \left(\frac{1}{\cos\left(T\right)}\right)}.\]
To get a fully thermodynamic decay, we must rescale the short-time regime $T$ such that $\frac{2}{T}\ln \left(\frac{1}{\cos\left(T\right)}\right)\approx \gamma$,
where $\gamma$ is a constant. This rescaling gives an exponential decay of the form:
\begin{equation}\label{expon}
\eta(t)=\frac{1}{2}\Big(\mathbb{I}^{\mathcal{S}} + e^{-\gamma (t-t_0)} a_j (t_0)\sigma_j^{\mathcal{S}} \Big).
\end{equation}
This evolution can be written as a first order differential equation,
$\dot{\eta}(t)= \gamma \left(\frac{1}{2}\mathbb{I}-\eta(t) \right)$.

If we choose $\mathrm{L}_0=\sqrt{\frac{\gamma}{3!}}\mathbb{I}$ and $\mathrm{L}_\alpha=\sqrt{\frac{\gamma}{3!}}\sigma_\alpha$ for $\alpha >0 $, the differential equation is of the form from Eq. (\ref{lindblad}).

\section{Non-Markovian Dynamical Maps}\label{sectnonmarko}

The Markov approximation is incompatible with a general theory of open quantum systems. In the previous section, we discussed how this approximation was taken to obtain an approximated composition property from Eq. (\ref{aproxcompo}) that permits the definition of the derivative of the map. By relaxing these assumptions, we can allow for not completely positive dynamical maps and thus account for physically meaningful correlations with the environment. We obtain the composition property by accounting for correlations.

\subsection{Initially Correlated States}

The tensor product of a system and environment state can be evolved to develop correlations. Their dynamical maps are computed as before, 
\begin{equation}\label{ mapdiag }
\begin{array}{ccc}
\rho(t_0)=\eta(t_0)\otimes\tau & \leftrightarrow & \rho(t_2)=\mathbb{U}_{(t_2|t_0 )}\Big( \rho(t_0) \Big)  \\ 
 \downarrow &   &   \downarrow \\
  \eta(t_0) & \rightarrow & \mathbb{B}_{(t_2|t_0 )}\Big(\eta(t_0)\Big)=\eta(t_2). \\
\end{array}
\end{equation}
To define the map $\mathbb{B}_{(t_2|t_0 )}$ we know the composition of three maps. First, one that inverts the trace at $t_0$, to have an arrow that goes from $\eta(t_0)\rightarrow \rho(t_0)$. Then we have a unitary map $\mathbb{U}_{(t_2|t_0 )}$. At time $t_2$ we have a trace to go from $\rho(t_0)\rightarrow\eta(t_0)$. In Eq. (\ref{ mapdiag }), to go from $\eta(t_0)\rightarrow\eta(t_2)$ we start on the bottom-left, go up, then to the right, and then down. If we introduce an intermediate time  $t_1$ the evolution becomes,
\begin{widetext} 
\begin{equation}\label{mapdiag2}
\begin{array}{ccccc}
\rho(t_0) & \leftrightarrow & \rho(t_1)=\mathbb{U}_{(t_1|t_0 )}\Big( \rho(t_0)\Big) & \leftrightarrow & \rho(t_2)=\mathbb{U}_{(t_2|t_1 )}\Big( \rho(t_1)\Big)\\ 
 \downarrow &   &   \downarrow &   &   \downarrow\\
  \eta(t_0) & \rightarrow & \mathbb{B}_{(t_1|t_0 )}\Big(\eta(t_0)\Big)=\eta(t_1) & \dashrightarrow & \mathbb{B}_{(t_2|t_1 )}\Big(\eta(t_1)\Big)=\eta(t_2).\\
\end{array}
\end{equation}
\end{widetext} 
$\mathbb{B}_{(t_2|t_0 )}$ as well as $\mathbb{B}_{(t_1|t_0 )}$ are completely positive, but $\mathbb{B}_{(t_2|t_1 )}$ might come from a $\rho(t_1)\neq \eta(t_1)\otimes\tau $. Not completely positive maps can come from system initial environmental correlations, such as entanglement \cite{pechukas94a} and more generalized quantum correlations \cite{Rodriguez07a}. 

To develop a prescription to consistently describe maps for initially correlated states, we need to find the inverse of the trace at time $t_1$, $\mathbb{T}\rho(t_1)=\eta(t_1)$, such that $\eta(t_1)\rightarrow\rho(t_1)$, use the inverse to find to dynamical map. Inverting the trace was accomplished in Section \ref{sectInitUnco} by introducing a completely positive embedding map, Eq.~(\ref{embedcp}). For initially correlated states it is necessary to relax the positivity condition. Complete positivity is a stronger condition than positivity, and will need to be relaxed as well, as was proposed by Pechukas \cite{pechukas94a}. We study how these not completely positive maps inverse maps have a physical interpretation if we account for non-Markovian quantum dynamics.

\subsection{Inverse Maps}\label{sectinve}

An inverse map connects the dynamics of the reduced state to the total state. With it, the composition property of the dynamical map can be found without approximations by exploiting the group property of the unitary maps $\mathbb{U}$ in the total space. The correlations that exists at time $t_1$  can be mapped back to a time $t_0$ when they were uncorrelated. Correlations at $t_1$, by definition, limit the valid domain of states at that time. Identically, the history from $[t_0,t_1]$ can limit the domain at time $t_1$. We treat correlations as a consequence of the memory effects from $[t_0,t_1]$. Non-Markovian dynamics are obtained from  system variables that are correlated with outside variables.

A consistent way to define maps after they have developed correlations is with inverse maps. Inverse maps have been studied before \cite{Jordan07b}, but here we consider the pseudo-inverse from Eq. (\ref{ainver}) 
and find a matrix inverse $\widetilde{\mathbb{A}}_{(t_i|t_f )}$ of the map $\mathbb{A}_{(t_f|t_i )}$ that evolves the state backwards in time. To define the map, additional information is necessary. This additional information is the history from the unitary evolution. From this, the inverse dynamical map $\widetilde{\mathbb{B}}_{(t_i|t_f )}$ can be found, which is generally not a positive map. $\widetilde{\mathbb{B}}_{(t_i|t_f )}$ can only be meaningfully applied to the set $\mathbb{B}_{(t_f|t_i )}\eta(t_i)$ for all density matrices $\left\{ \eta(t_i) \right\}$. The compatibility domain is  identical to the set of states compatible with the history from $[t_0,t_1]$. States outside the compatibility domain will be inconsistent with its history, and when its evolution is reversed it may not be mapped to a valid physical state. There is no reason for these maps to be positive, much less completely positive. On the contrary, history effects create correlations that limit the domain of validity. 

Experimentally, inverse maps can be found from sufficient knowledge of their forward counterparts. Since we are considering finite-dimensional environments, the evolution will have Poincar\'{e} recurrences in it. The recurrence time gets longer as the environment gets larger. The evolution of a system state of $N$ dimensions can always be modelled with an environment with $N^2$ dimensions \cite{sudarshan86}. This makes the number of parameters finite and the problem tractable. A related way to determine the inverse maps to know its forward counter part to high enough orders in time. Such as scheme for a one-qubit system coupled to a one-qubit environment was developed in \cite{Jordan07c}. There a finite number of derivatives at the initial time almost fully characterize the evolution of the total state.  Another procedure for determining the total Hamiltonian of qubit systems is given in \cite{Mohseni07a}.

In realistic circumstances, full knowledge of the map may not be accomplished. In Section \ref{sectnoneq} we will show how incomplete knowledge of the evolution leads to irreversibility. Now, we use the inverse map to derive the canonical dynamical map that accounts for correlations with the environment.

\subsection{Canonical Dynamical Map}

With the inverse map $\widetilde{\mathbb{B}}$, we can now define a \emph{canonical dynamical map} $\mathbb{B}^\mathsf{C}$ for states win initially environmental correlations. The knowledge necessary for finding the inverse map are the additional variables needed to extend a system space of a non-Markovian evolution. 

\begin{figure}[htb]
\resizebox{9 cm}{4.5 cm}{\includegraphics{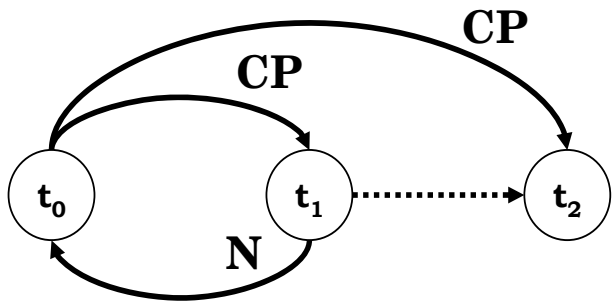}}
\caption{   This diagram represents the evolution described by Eq.~(\ref{mapdiag2}).  \textbf{CP} is Completely Positive evolution, \textbf{N} is Not Positive Evolution. A canonical dynamical map from $t_1 \rightarrow t_2$ can be defined going to $t_0$, and from there forward to $t_2$, as in Eq.~(\ref{canon}).}
\label{Figu1}
\end{figure}

We compose the maps to find the evolution described in Eq. (\ref{mapdiag2}) from $t_1\rightarrow t_2$ as in Fig. (\ref{Figu1}).
 First, we map the state from $t_1$ to $t_0$ using the inverse map, then evolve the state forward to $t_2$. We write this as
\begin{eqnarray}\label{canon}
\mathbb{B}^\mathsf{C}_{(t_2|t_1 )} \equiv \mathbb{B}_{(t_2|t_0 )}\star \widetilde{\mathbb{B}}_{(t_0|t_1 )}.
\end{eqnarray} 
The composition is easily computed in the $\mathbb{A}$ form of the map. From Eq.~(\ref{canon}), the canonical dynamical maps have the composition property: 
\begin{eqnarray}\label{canoncompo}
\mathbb{B}^\mathsf{C}_{(t_f|t_i )} = \mathbb{B}^\mathsf{C}_{(t_f|t )}\star \mathbb{B}^\mathsf{C}_{(t|t_i )},
\end{eqnarray} without need of any approximations.

It has been implied that $t_0 < t_1 < t_2$, but this needs not be. If $t_1=t_0$, the original completely positive map is obtained. If $t_0\leq t_1$ but $t_2=t_0$, we obtain
\begin{eqnarray}\label{canoninver}
\mathbb{B}^\mathsf{C}_{(t_0|t_1 )} = \mathbb{B}_{(t_0|t_0 )}\star \widetilde{\mathbb{B}}_{(t_0|t_1 )}=\widetilde{\mathbb{B}}_{(t_0|t_1 )},
\end{eqnarray}
using $\mathbb{B}_{(t_0|t_0 )}=\mathbb{I}$, where $\mathbb{I}$ is the identity map. Since $\mathbb{B}^\mathsf{C}_{(t_i|t_f )}\star \mathbb{B}^\mathsf{C}_{(t_f|t_i )}=\mathbb{B}^\mathsf{C}_{(t_f|t_i )}\star \mathbb{B}^\mathsf{C}_{(t_i|t_f )}=\mathbb{I}$, we conclude that inverse maps are canonical maps. Canonical maps have the composition property from Eq.~(\ref{canoncompo}) and have an inverse from Eq.~(\ref{canoninver}), forming a one parameter group in time. They preserve the trace and Hermiticity, but they are in general not positive and are only valid within their compatibility domain. This is what we wanted: a map that allows for correlations with the environment such that any incompatible state with the correlations will give an unphysical total state. If we had full knowledge of the time dependence of the canonical dynamical map, it would be fully irreversible. Only some canonical maps $\mathbb{B}^\mathsf{C}_{(t^\prime|t )}$ (the unitary map) might be completely positive for any choice of $t$ and $t^\prime$. 
The derivative of the canonical dynamical map is also well defined for all times. 

The canonical dynamical map we have defined describe the most general dynamics of an open quantum systems with out the need for the Markov approximation. In the next part, we show the connection between the total unitary dynamics and the correlations with the environment to the reduced dynamics given by the cannonical dynamical map.

\subsection{Canonical Embedding Map}\label{sectembed}
 The canonical dynamical map is connected to the reduced evolution of the system and the environment. To go from the reduced system state, to the total evolution we need to invert the trace at all times. We had used an embedding map $\mathbb{E}_{t_i}$ that could consistently invert the trace map $\mathbb{T}$ for states uncorrelated at time $t_i$. For clarity, we will focus only on the embedding from Eq.~(\ref{embedcp}) for initially uncorrelated states. Even so, this procedure also work for any valid embedding (completely positive or not completely positive) such as the ones proposed by Pechukas \cite{pechukas94a} and Alicki \cite{alicki95}. With the use of the canonical map, we can generalize these embedding maps to all times, even when the initial correlations have evolved, such that: 
\begin{equation}
\eta(t)\rightarrow\mathbb{E}^\mathsf{C}_{t }\,\eta(t)=\rho(t) \mbox{ for all } t.\nonumber
\end{equation}
Such an embedding map will use the history of the reduced evolution to ``close'' the open system. We evolve the state backwards to the time where we had defined a valid embedding map, undo the trace, and unitarily go forward. From Eq.~(\ref{DynMapPicture}), this would be pictorically represented by:
 \begin{equation}\nonumber
\begin{array}{cccc}
&   &  \mathbb{U}_{(t|t_0 )}  & \\
&  \rho(t_0) & \Rightarrow & \rho(t)=\mathbb{E}^\mathsf{C}_{t }\eta(t) \\ 
 &\mathbb{E}_{t_0 } \; \;  \;  \;  \Uparrow \; \; \; \;   \; & & \uparrow  \\
&  \eta(t_0) & \Leftarrow & \eta(t). \\
&   &  \mathbb{B}^\mathsf{C}_{(t_0|t )}  &
\end{array}
\end{equation}
This \emph{canonical embedding map} from $\eta(t)\rightarrow\rho(t)$ is defined as:
\begin{eqnarray}\label{embedmap}
\mathbb{E}^\mathsf{C}_{t } \equiv  \mathbb{U}_{(t|t_0 )} \star \mathbb{E}_{t_0 }\star \mathbb{B}^\mathsf{C}_{(t_0|t )}.
\end{eqnarray}

The canonical embedding map preserves Hermiticity and trace, but might not be positive. Its compatibility domain corresponding to the system space compatible with the correlations existing at time $t$. The set of states that will give unphysical evolutions is also incompatible with the memory effects of the environment. 

We do not need an embedding map for an uncorrelated total state at $t_0$; any valid embedding for any other time $t$ will do. This is proven by:
\begin{eqnarray}\label{embedmap2}
\mathbb{E}^\mathsf{C}_{t } &=&  \mathbb{U}_{(t|t_0 )} \star \mathbb{E}_{t_0 }\star \mathbb{B}^\mathsf{C}_{(t_0|t )},\nonumber\\
&=&  \mathbb{U}_{(t|t^\prime )} \star \Big( \mathbb{U}_{(t^\prime|t_0 )} \star \mathbb{E}_{t_0 }\star \mathbb{B}^\mathsf{C}_{(t_0|t^\prime )} \Big) \star \mathbb{B}^\mathsf{C}_{(t^\prime|t )}, \nonumber\\ &=&\mathbb{U}_{(t|t^\prime )} \star \mathbb{E}_{t^\prime }\star \mathbb{B}^\mathsf{C}_{(t^\prime|t)}.
\end{eqnarray}
By knowing one embedding map for a time $t^\prime$ and the unitary operator in the interval $[t,t^\prime]$, any other embedding for another $t$ can be found.

This approach explicitly shows the connection between the correlations of the state with the environment and its history. Correlations at one time can be changed to correlations at another as long as the history is known. The necessity of additional knowledge to establish an embedding map makes it non-Markovian. The possible negativity of the map shows that the history limits some of the states in the system space to be compatible with the total system-environment state. Detailed knowledge of the reduced dynamics can be used to find the full dynamics \cite{Jordan07c,Mohseni07a}. The embedding map will be used as a mathematical device to connect the full dynamics to those of the reduced space without need of the Markov approximation.

\subsection{Example}\label{examp2}

We return to the example from Section \ref{examp1} to illustrate how to compute an inverse map, then the canonical dynamical map and the embedding map. We want to map the Bloch vector $\overrightarrow{a}$ from the final time $t_f$ to the initial time $t_i$. In its affine form this is:$
\overrightarrow{a}(t_i)=\overline{R}^{-1}_{(t_f|t_i )}\cdot\left(\overrightarrow{a}(t_f)-\overrightarrow{r}\right)$.
For the particular example from Eq. (\ref{exampAffine1}), $\overrightarrow{a}(0)=\frac{1}{c^2}\overrightarrow{a}(t)$. The inverse $\widetilde{\mathbb{A}}_{(t_0|t )}$ can be found from the dynamics,
\begin{eqnarray}
\widetilde{\mathbb{A}}_{(t_0|t )}=\frac{1}{2}\left(%
\begin{array}{cccc}
  1+c^{-2} & 0 & 0 & 1-c^{-2} \\
  0 & 2 c^{-2} & 0 & 0 \\
  0 & 0 & 2c^{-2} & 0 \\
  1-c^{-2} & 0 & 0 & 1+c^{-2} \\
\end{array}%
\right). \nonumber
\end{eqnarray}By index exchange, we obtain $\widetilde{\mathbb{B}}_{(t_0 |t)}$, which in the eigen-system representation is:
\begin{eqnarray}\label{exampinverse1}
 \lambda_0 (t-t_0)=\frac{1}{2}\left(1+3c^{-2} \right),&\mathrm{C}_0=\frac{1}{\sqrt{2}}\openone,\nonumber\\
\lambda_{1,2,3} (t-t_0)=\frac{1}{2}\left(1-c^{-2} \right),&\mathrm{C}_{1,2,3}=\frac{1}{\sqrt{2}}\sigma_{1,2,3}.
\end{eqnarray} 
For certain values of $t$, $\widetilde{\mathbb{B}}$ is not completely positive. This represents the periodic behavior of the original map: as the state is squeezed, the compatibility domain of its inverse maps also shrinks. For the times where $c=0$, the only compatible state is the center of the Bloch sphere. States outside the compatibility domain are not relevant to the physical dynamics of the open system. They are inconsistent with the developed correlations and history. 

We can define the canonical dynamical map by means of Eq.~(\ref{canon}). The composition property is easier to apply on the $\mathbb{A}$ form of the map, since it is matrix multiplication. We compute $\mathbb{A}_{(t^\prime|t_0 )}\cdot\widetilde{\mathbb{A}}_{(t_0|t)}=\mathbb{A}_{(t^\prime|t )}$, then exchange the indices to obtain the $\mathbb{B}^\mathsf{C}$ form of the canonical map, that has as its eigensystem:
\begin{eqnarray}
 \lambda_0 (t^\prime-t)=\frac{1}{2}\left(1+3\frac{c^{2}}{\widetilde{c}^{2}} \right),&\;&\mathrm{C}_0=\frac{1}{\sqrt{2}}\openone,\nonumber\\
\lambda_{1,2,3} (t^\prime-t)=\frac{1}{2}\left(1-\frac{c^{2}}{\widetilde{c}^{2}}  \right),&\;&\mathrm{C}_{1,2,3}=\frac{1}{\sqrt{2}}\sigma_{1,2,3},\nonumber
\end{eqnarray} 
where $c\equiv \cos(t^\prime -t_0)$ and $\widetilde{c}\equiv \cos(t-t_0)$. With $t=t_0$ the map is completely positive. Taking $t^\prime=t$ gives the inverse map.

Finally, a canonical embedding map can be computed from Eqs.~(\ref{embedcp}), (\ref{embedmap}) and (\ref{exampinverse1}):
\begin{eqnarray}\label{ex1embedmap}
\mathbb{E}^\mathsf{C}_{t }\Big( \eta(t) \Big)=U_{(t|t_0 )}\left(\left[ \mathbb{B}^\mathsf{C}_{(t_0|t )}\Big( \eta(t)\Big)\right] \otimes \tau \right)U_{(t|t_0 )}^\dagger.\nonumber
\end{eqnarray}
From Eq.~(\ref{examptensor1}), with $\eta(t)=\frac{1}{2}\left(\openone + a_j(t) \sigma_j \right)$ and $\tau=\frac{1}{2}\openone$, we carry out the calculation to reach the final result:
\begin{eqnarray}\label{ex2embedmap}
\mathbb{E}^\mathsf{C}_{t }\Big( \eta(t) \Big)=\frac{1}{4}\left[ \openone \otimes \openone+ a_j(t)\Big(\sigma_j\otimes \openone \right.\nonumber\\ \left.+ \tan(t)^2 \openone\otimes\sigma_j +\tan(t) \left( \sigma_k\otimes\sigma_l-\sigma_l\otimes\sigma_k \right)  \Big)\right],
\end{eqnarray} summing over index $j$, with $\left\{ j,k,l \right\}$ being cyclic. The compatibility domain is represented here by the unbounded character of $\tan(t)$. Periodically the compatible set of vectors $\overrightarrow{a}(t)$ tend to the center of the Bloch sphere. The compatible system parameters change periodically with the correlations.

In the next section, we show how the derivative of the canonical dynamical map is related to the embedding map.

\section{Non-Markovian master equation}\label{sectmaster}

The non-Markovian master equation can be derived from the canonical dynamical map from Eq.~(\ref{canon}). The time derivative of the unitary operator is $\dot{U}=-iH U$, so it follows that the time derivative of the canonical map is:
\begin{eqnarray} \frac{\partial}{\partial t}\mathbb{B}^\mathsf{C}_{(t|t_i )}\Big(\eta(t_i)\Big)
=-i \mbox{Tr}_{\mathcal{E}}\left[H U_{(t|t_i )}
\rho(t_i) U^\dag_{(t|t_i )}\right]\nonumber\\+i \mbox{Tr}_{\mathcal{E}}\left[U_{(t|t_i )}
\rho(t_i) U^\dag_{(t|t_i )}H\right].\nonumber
\end{eqnarray} 
This is equivalent to a von Neumann equation  of the reduced system space, $\mbox{Tr}_{\mathcal{E}}\left[ \dot{\rho}(t)\right]=-i\mbox{Tr}_{\mathcal{E}}\left[ H,\rho(t)\right]$. To show how the differential equation can explicitly depend only on the system space, we use the embedding map $\mathbb{E}^\mathsf{C}_{t }$ from Eq.~(\ref{embedmap}). The differential equation is:
\begin{equation}\label{Master2} 
\frac{\partial }{\partial t}\eta(t)=-i\mbox{Tr}_{\mathcal{E}}\left[ H,\mathbb{E}^\mathsf{C}_t\Big(\eta(t)\Big)\right].\nonumber
\end{equation}

Now, we write the total Hamiltonian  as $H=H_O+H_I$, where $H_O$ is the local (system) part of the Hamiltonian. This local part acts through the embedding map leaving it unchanged. With this, we have the standard form of the non-Markovian master equation:

\begin{equation}\label{Master3} 
\frac{\partial }{\partial t}\eta(t)=-i\Big[ H_O,\eta(t)\Big]+\mathbb{K}_t\Big( \eta(t)\Big),
\end{equation} 
with $\mathbb{K}_t(\cdot)\equiv\mathbb{F}_t(\cdot)+\mathbb{F}^\dagger_t(\cdot)$, where
\begin{eqnarray}\label{GOper}
\mathbb{F}_t\left( \cdot \right) &=&- i\mbox{Tr}_{ \mathcal{E}}\left[ H_I \mathbb{E}^\mathsf{C}_t( \cdot ) \right],\\
\mathbb{F}^\dagger_t\left( \cdot \right) &=&+i\mbox{Tr}_{ \mathcal{E}}\left[ \mathbb{E}^\mathsf{C}_t( \cdot )H_I  \right].\nonumber
\end{eqnarray} 
The embedding map here is just a mathematical device that allow us to show that the
Hermitian super-operator $\mathbb{K}_t$ is related to the time derivative of the canonical dynamical map by:
\begin{equation}\label{Master4}
\frac{\partial }{\partial t}\mathbb{B}^\mathsf{C}(\cdot)=-i\left[H_O,\cdot  \right]+\mathbb{K}_t(\cdot).
\end{equation}
The $H_O$ term is the Hamiltonian evolution of the system and $\mathbb{K}_t$ carries all the effects of the environment, including dissipation and memory. 

Since the environment is finite-dimensional, there will be some quasi-periodicity to this evolution as information goes from the system to the environment, and back. At certain times the space is being contracted, while at others it is expanded. These Poincar\'{e} recurrences are a consequence of the canonical maps forming a group. This should be contrasted to the Kossakowski-Lindblad master equation, that uses the Markov approximation to obtain a dynamical semigroup. The Markovian master equation can be obtained by rescaling $\mathbb{K}_t$ to be time independent.

The non-Markovian master equation is generalization of the von Neumann equation to open quantum systems.

\subsection{Example}\label{ex1master}

We will complete the example from Section \ref{examp2} to illustrate the consistency of Eq.~(\ref{Master4}). In this case, $H_O=0$, $H_I=\frac{1}{2}\sum_j \sigma_j\otimes\sigma_j$ and $\mathbb{E}^\mathsf{C}_t( \eta(t) )$ was calculated in Eq.~(\ref{ex2embedmap}). In this case, $\mathbb{F}_t$ from Eq.~(\ref{GOper}) becomes\begin{eqnarray}
\mathbb{F}_t\Big( \eta(t) \Big) = \sum_j \frac{1}{4} \Big( -i\tan(t)^2-2\tan(t)  \Big) a_j (t) \sigma_j.\nonumber
\end{eqnarray}
The non-Markovian master equation is then:
\begin{eqnarray}\label{ex1masterG}
\dot{\eta}(t)=\mathbb{K}_t\Big( \eta(t) \Big) = -\sum_j \tan(t)a_j(t)\sigma_j,\nonumber\\=2\tan(t)\Big(\openone-2\eta(t)\Big).
\end{eqnarray}
If we only look at the $\sigma_j$ component, the evolution of its expectation value is:
\begin{equation}\label{ex1difeq}
\dot{a_j}(t)=-2 \tan(t)a_j(t),\nonumber
\end{equation}
and has as solution $a_j(t)=\cos(t-t_0)^2 a_j(t_0)$. This which agrees with the starting point from Eq.~(\ref{exampAffine1}). 

This is an example of how the total evolution and the non-Markovian dynamical map are related to each other. Similarly, if we know the first derivative of the evolution, as in Eq. (\ref{ex1masterG}), partial knowledge of the multiplication of $H_I$ and $\mathbb{K}_t$ can be determined from Eq. (\ref{GOper}). Higher derivatives of the canonical map may yield even more information of the full dynamics.

Note that there is no dissipation in this equation because we allow for Poincar\'{e} recurrences. In the next section we discuss how partial knowledge of the evolution leads to decay beyond thermodynamic equilibrium.

\section{Non-Equilibrium Quantum Thermodynamics}\label{sectnoneq}

The Kossakowski-Lindblad master equation for a process may be obtained by taking the Markov approximation of the dynamical map of that process. From this approximation, irreversibility is introduced and relaxation into thermodynamic equilibrium obtained. Exponential decays are natural solutions to many instances of this equation.

However, the non-Markovian master equation from Eq.~(\ref{Master4}) allows us to know the full evolution of the system without irreversibility. Thermodynamic effects can be introduced by expanding $\mathbb{K}_t$ for short times without the need of the Markov approximation. As larger orders in time are introduced to the approximation, longer memory effects and higher order correlations with the environment appear. Higher orders in time allow us to go beyond the thermodynamic regime; non-equilibrium quantum thermodynamical effects can be studied. We illustrate this with an example.

\subsection{Example}

The master equation from the example in Section \ref{ex1master} is not only non-Markovian, it is also periodic. To introduce some dissipation and decay, and connect it to non-equilibrium thermodynamics, we must make an approximation for short times in the master equation. Now only memory effects of a small order in time will be kept as the approximation discards some knowledge of the evolution. Irreversibility arised from the limited information.  

Experimentally, limited information comes from monitoring the system for only time shorter that Poincar\'{e} time, and trying to find the master equation from this incomplete information. It can also come from knowing the dynamics of the reduced system to a limited number of derivatives only.

We approximate $\tan(t)\approx t$ and Eq.~(\ref{ex1masterG}) becomes:
\begin{eqnarray}
\dot{\eta}(t)=2t\Big(\openone-2\eta(t)\Big).\nonumber
\end{eqnarray} The evolution of just one component is,
$\dot{a_j}(t)=-2 t \;a_j(t).$
The solution to this differential equation is 
\begin{eqnarray}\label{gausdecay}
a_j(t-t_0)=e^{-(t-t_0)^2}a_j(t_0).\end{eqnarray}  As time goes to infinity, the polarization of the Bloch vector shrinks to zero through a non-exponential decay due to the short-time memory effects retained from the environment. In other words, the environment is not an ideal (passive) thermodynamic bath as it is dynamically allowed to react slightly. This is an example of a non-equilibrium quantum thermodynamical effect. The decay of the form $e^{-t^2}$ from Eq. (\ref{gausdecay}) should be contrasted to the thermodynamic decay $e^{-\gamma t}$ from Eq.~(\ref{expon}). The non-equilibrium thermodynamic decay can be faster than exponential for very small values of $\gamma$, while it can be slower for large values of $\gamma$. At intermediate values, $\gamma \approx 1$, the non-Markovian decay is slower than exponential at first, and then much faster. Accounting for memory effects can make decays faster or slower.

The non-Markovian decay also differs from exponential decay close to the initial time. In this non-equilibrium thermodynamic solution, the initial time derivative of the polarization is zero, which is crucial to obtaining the quantum Zeno effect \cite{Misra77a}. Before, quantum Zeno could be obtained only from the Hamiltonian part of the Kossakowski-Lindblad master equation. Now, even the interaction with the environment can give rise to a Zeno region.

\section{Discussion and Conclusions }\label{sectconclu}

We have developed a generalized non-Markovian master equation for open quantum systems by accounting for correlations with the environment. Previous work on completely positive non-Markovian master equations can be treated as special classes of the non-markovian master equation in this paper. For example,  Shabani and Lidar proposed a class of master equations whose memory comes from total states with correlations derived from measurement approach \cite{Shabani05a}. This is equivalent to having an embedding map from  Eq.~(\ref{embedmap2}) for the particular time $t^\prime$ given by a measurement on the environment. From this, a canonical embedding equation can be developed for all times, and their master equation obtained. This class of embedding is completely positive, at the expense of limiting to only classical correlations of the environment with the system at time $t^\prime$ \cite{Ollivier01a,Rodriguez07a}. Breuer proposes another class of embedding maps for a different restricted kind of correlations that come from the projection-operator method \cite{Breuer07a}. Our approach permits any kind of correlations, classical or quantum.

In conclusion, we have discussed how not completely positive dynamical maps in open quantum systems represent the limited domain due to correlations with the environment.  With this, a canonical dynamical map was developed that can be applied for any initially correlated systems. The canonical dynamical maps form a dynamical group, different from the dynamical semigroup from the Kossakowski-Lindblad equation. A canonical embedding map can be constructed to express the correlations with the environment at any time, effectively closing the evolution of the open system.  A generalized non-Markovian master equation was constructed that was local in time and corresponds to the reduced-space von Neumann equation. Approximations to this equation, such as the ones given by a limited knowledge of the history, or knowledge of the evolution of the system to a small order in time, can lead to irreversible behavior beyond the purely thermodynamic regime. This theory permits the study of non-equilibrium quantum thermodynamic effects.

We would like to thank A. Shaji, T. Tilma, K. Modi and M. Mohseni for insightful discussions. We are also grateful to A. Chimonidou and K. Dixit for their comments and corrections to the manuscript, and D. Miracle for massive help with the editing.

\vspace{4 mm}

\appendix

\section{Classical Stochastic Processes}\label{classtoc}

A classical probability vector $\overrightarrow{p}(i)$ can be evolved into another one, $\overrightarrow{p}(f)$, by means of a matrix $\mathbb{M}$ using the equation,  \[\overrightarrow{p}(f)=\mathbb{M}\cdot \overrightarrow{p}(i).\] The probability vectors form a convex set; for a finite number of nonzero components it is a simplex. If the vectors are written in tensor notation, the evolution is fully determined by $p(f)_{r^\prime}=\mathbb{M}_{r^\prime ,r}p(i)_r$. If $\mathbb{M}$ is treated as a map, it must have as its domain all probability vectors $\{ \overrightarrow{p}(i) \}$, and as its image a subset of the domain. Matrices with these properties are called stochastic matrices. The only stochastic maps that are invertible for the whole domain are the permutations of the vertices of the simplex. These stochastic matrices, whose inverse happen to be also a stochastic matrix, correspond to maps whose domain and image are the whole set $\{ \overrightarrow{p} \}$. They form a special subclass called bi-stochastic matrices.  If an inverse is desired for more general cases, caution must be taken on where it acts. The pseudo-inverse of a stochastic matrix $\widetilde{\mathbb{M}}$, such that $\widetilde{\mathbb{M}}\cdot \mathbb{M} =\mathbb{I}$ might itself not be a stochastic matrix. $\widetilde{\mathbb{M}}$ is properly defined only on the subset of probability vectors of the form $\mathbb{M}\cdot \overrightarrow{p}$ for all $\{ \overrightarrow{p} \} $.

The probability vectors can be evolved as a process in time with a stochastic map, $\overrightarrow{p}(t_f)=\mathbb{M}_{(t_f|t_i)}\cdot\overrightarrow{p}(t_i)$. If $\overrightarrow{p}(t_f)$ depends only on the particular state $\overrightarrow{p}(t_i)$, it is said to be a Markov process. Markovian processes correspond to the loss of information in a mononotonic fashion. Sometimes a process $\mathbb{N}$ for a time interval $[t_i\rightarrow t_f]$ must be defined using outside variables $\overrightarrow{r}(t_i)$. If so, the process is described by $\overrightarrow{q}(t_f)=\mathbb{N}_{(t_f|t_i)}\left( \overrightarrow{r}(t_i)\right)\cdot\overrightarrow{q}(t_i)$ and is said to be non-Markovian. The additional variables $\overrightarrow{r}$ could represent the state $\overrightarrow{q}(t)$ at other times $t\neq t_i$, and may be referred to as memory effects or history. In such a case, the knowledge of $\overrightarrow{r}$ is the history needed to consistently define $\mathbb{N}$. 

A Markovian process in $\overrightarrow{p}$ can be made non-Markovian in $\overrightarrow{q}$ by reducing the space of known parameters:
\begin{eqnarray}\label{NMvsM} 
&\mbox{\emph{Markovian:}}&  \overrightarrow{p}(t_f)  =  \mathbb{M}_{(t_f|t_i)} \cdot \overrightarrow{p}(t_i) \nonumber \\
 & \downarrow & \nonumber \\ 
& \mbox{\emph{Non-Markovian:}}&   \overrightarrow{q}(t_f)  =  \mathbb{N}_{(t_f|t_i)}\Big( \overrightarrow{r}(t_i) \Big)\cdot\overrightarrow{q}(t_i). \nonumber
\end{eqnarray}
On the other hand, this non-Markovian process $\mathbb{N}$ may be mapped into a Markovian process $\mathbb{M}$ by extending the space from $\overrightarrow{q}(t_i)$ to $\overrightarrow{p}(t_i) \equiv \left\{\overrightarrow{q}(t_i),\overrightarrow{r}(t_i) \right\}$.  This would require additional knowledge of the process. In general, $\overrightarrow{q}$ are correlated to $\overrightarrow{r}$. Not all $\overrightarrow{q}$ are permitted; they must be compatible with their corresponding $\overrightarrow{r}$.

A physical example is the process of classical scattering of multiple particles. This is a Markovian evolution in position  and momentum variables; if integrated over momentum variables, correlations with it are now folded into a memory kernel that leads to non-Markovian effects. A method that extends the space to get from non-Markovian phenomena to Markovian by studying the time evolution of the position variables was computed for scattering particles in \cite{Lavakore62a}. 

\section{Derivation of the Kossakowski-Lindblad master equation from the dynamical map}\label{dkossak}

We derive the Kossakowski-Lindblad master equation from the dynamical map following the procedure given in 
\cite{shaji05dis} and using the assumptions and approximations discussed in Section {\ref{kossak}. The Kossakowski-Lindblad master equation was derived originally for completely positive evolutions, such that $0\leq \lambda_\alpha$ in Eq. (\ref{CMap}). With it, we can write:
\begin{eqnarray}
\sqrt{\lambda_0}\mathrm{C}_0 &\approx&\openone+\sqrt{t}\;\mathrm{L}_0, \nonumber \\
\sqrt{\lambda_\alpha}\mathrm{C}_\alpha &\approx& \sqrt{t}\;\mathrm{L}_\alpha \quad \mbox{for} \quad \alpha >0.
\end{eqnarray}
The action of the map for small $t$ is:
\begin{eqnarray}\label{deriv01}
\eta(t)&=&\Big( \openone+\sqrt{t}\;\mathrm{L}_0 \Big)\; \eta(0) \;\big(\openone+\sqrt{t}\;\mathrm{L}_0 \big)^\dagger \nonumber\\ &+&\sum_{\alpha>0} t\;\mathrm{L}_\alpha \; \eta(0) \; \mathrm{L}_\alpha^\dagger.  
\end{eqnarray}
Using the trace preservation condition $\sum_\alpha \lambda_\alpha \mathrm{C}_\alpha^\dagger \mathrm{C}_\alpha=\openone$ we find the property:
\begin{eqnarray}
\frac{1}{\sqrt{t}}\big( \mathrm{L}_0^\dagger + \mathrm{L}_0 \big)=-\sum_{\alpha}\;\mathrm{L}_\alpha^\dagger \mathrm{L}_\alpha.\nonumber
\end{eqnarray}
With this, Eq.~(\ref{deriv01}) can be rewritten as:
\begin{eqnarray}\label{deriv02}
\frac{\eta(t)-\eta(0)}{t}=\frac{1}{2\sqrt{t}}\Big[\mathrm{L}_0-\mathrm{L}_0^\dagger,\;  \eta(0) \Big]\nonumber\\+\frac{1}{2}\sum_\alpha\Big( \big[\mathrm{L}_\alpha\; \eta(0),\mathrm{L}_\alpha^ \dagger\big]+\big[\mathrm{L}_\alpha, \eta(0)\;\mathrm{L}_\alpha^ \dagger\big]\Big).
\end{eqnarray}
Looking at Eq.~(\ref{deriv02}) we note that $\mathrm{L}_0-\mathrm{L}_0^\dagger$ is anti-Hermitian and could be written as $\mathrm{L}_0-\mathrm{L}_0^\dagger\equiv -i\sqrt{t}H$. The Hermitian operator $H$ has now been rescaled by a factor of $\sqrt{t}$ to implicitly carry a time dependence. With this, and taking the limit of Eq.~(\ref{deriv02}) where $t\rightarrow 0$, the Kossakowski-Lindblad master equation from Eq. (\ref{lindblad}) is obtained. It corresponds to a Markovian process where $H$ is the (rescaled) effective local evolution, resembling a Hamiltonian, while $\mathrm{L}_\alpha$ are the operators that generate the completely positive dynamical semigroup \cite{Kossakowski72a,Gorini76a,Lindblad76a}. Dynamical semigroups\index{dynamical semigroup} are not groups as they do not have an inverse. Their irreversibility are consequence of all the assumptions and approximations.

\bibliography{refs2008}
\end{document}